# Gender-Based Heterogeneity in Youth Privacy-Protective Behavior for Smart Voice Assistants: Evidence from Multigroup PLS-SEM

Molly Campbell
*Computer Science Department*
*Vancouver Island University*
Nanaimo, Canada
molly.campbell@viu.ca

Yulia Bobkova
*Computer Science Department*
*Vancouver Island University*
Nanaimo, Canada
yulia.bobkova@viu.ca

Ajay Kumar Shrestha
*Computer Science Department*
*Vancouver Island University*
Nanaimo, Canada
ajay.shrestha@viu.ca

***Abstract*— This paper investigates how gender shapes privacy decision-making in youth smart voice assistant (SVA) ecosystems. Using survey data from 469 Canadian youths aged 16–24, we apply multigroup Partial Least Squares Structural Equation Modeling to compare males (N=241) and females (N=174) (total N = 415) across five privacy constructs: Perceived Privacy Risks (PPR), Perceived Privacy Benefits (PPBf), Algorithmic Transparency and Trust (ATT), Privacy Self-Efficacy (PSE), and Privacy Protective Behavior (PPB). Results provide exploratory evidence of gender heterogeneity in selected pathways. The direct effect of PPR on PPB is stronger for males (Male: $\beta = 0.424$; Female: $\beta = 0.233$; $p < 0.1$), while the indirect effect of ATT on PPB via PSE is stronger for females (Female: $\beta = 0.229$; Male: $\beta = 0.132$; $p < 0.1$). Descriptive analysis of non-binary (N=15) and prefer-not-to-say participants (N=39) shows lower trust and higher perceived risk than the binary groups, motivating future work with adequately powered gender-diverse samples. Overall, the findings provide exploratory evidence that gender may moderate key privacy pathways, supporting more responsive transparency and control interventions for youth SVA use.**

## I. Introduction

Smart voice assistants (SVAs) are increasingly integrated into the daily lives of youth, with technology that occupies both private and public spaces [1]. The always-on, voice-activated nature of these devices creates unique privacy challenges, where conversations and ambient data can be continuously captured and stored [2], [3], [4]. As a result, youth must navigate complex privacy dynamics with limited transparency about data collection and use.

These dynamics are not experienced uniformly. Gender is a key determinant that shapes how individuals perceive risk, build trust, and behave in digital environments. A growing body of research has documented gender differences in privacy attitudes and behavior in online and mobile contexts. Studies consistently show that women express higher privacy concerns and engage in more frequent protective behaviors than men, a pattern often linked to gendered socialization [5], [6]. The pathways to protection have also been observed to be gendered, with gender moderating key relationships like self-efficacy and threat avoidance [7]. This landscape is complicated by the privacy paradox, the gap between concern and action, which manifests differently by gender [8].

Despite this literature, relatively little is known about whether and how gender shapes privacy pathways specifically in youth SVA ecosystems. Prior work on SVAs examined general privacy perceptions and behaviors in adult populations [2] or has focused on gendered design and interaction dynamics [9], [10]. However, these studies have not examined the constructs central to privacy decision making, such as self-efficacy, trust, risk, and protective behavior, to test for gendered differences in the relationships between them. Our study addresses the gap by investigating whether established pathways to privacy protection are gendered for young SVA users.

This study builds on prior work on youth privacy in SVA use. The foundation was established through qualitative analysis of focus groups [11] and formalized into a quantitative structural model. The model was first validated on a combined youth sample [12] and subsequently used to investigate age-differentiated pathways [13]. The present study expands on this research by applying the same validated model to examine gender-differentiated pathways. A separate publication is warranted because the gender-based multi-group analysis (MGA) introduces additional analytical complexity, requiring focused interpretation of subgroup-specific path differences and their theoretical and practical implications. By conducting MGA by gender, this study moves beyond the baseline model to examine how gender shapes privacy decision-making. This study is guided by three research questions (RQs):

- RQ1: Do participants who self-report as female versus male differ in their average levels of PPR, PPBf, ATT, PSE, and PPB?
- RQ2: Do the structural relationships among these constructs, particularly the pathway from privacy self-efficacy to privacy-protective behavior, differ between the female and male groups?
- RQ3: Descriptively, how do participants who self-report as non-binary/other or select "prefer not to say" compare on these constructs relative to the female and male groups?

The remainder of this paper is organized as follows: Section II provides background and related works. Section III holds the methodology. Section IV presents the results of our study. Section V provides the discussion. Finally, section VI concludes the paper.

## II. BACKGROUND

### A. Gender and Digital Privacy Attitudes and Behavior

Prior work in privacy and human-computer interaction has repeatedly documented gender differences in digital privacy perceptions and behaviors. Evidence suggests that women and girls report higher privacy concerns and engage in more frequent privacy-protective behaviors than men and boys [5]. This pattern, emerging in adolescence, is linked to gendered socialization around risk and safety, where girls are often encouraged to be more cautious and threat-aware [6], [14]. Research indicates that gender moderates the psychological pathways to protective action. Studies on technology threat avoidance find that gender influences how self-efficacy and perceived threat relate to protective motivation and behavior [7]. Research further reveals differing decision-making processes, for example, in location sharing contexts, benefit is a primary driver for men, while privacy risks are a stronger deterrent for women [15]. These patterns can be shaped early, as gendered approaches to parental mediation contribute to differences in adolescents' privacy self-efficacy [16]. The privacy paradox, an observed digital phenomenon where privacy concern does not lead to protective action, adds complexity. However, research suggests this paradox may be moderated by gender, interacting with established differences in self-efficacy and risk perception [7].

### B. Smart Voice Assistants and Gender

In the context of smart voice assistants, gender may shape both experience and risk. Commercial SVAs are often shipped with default "female" voices and are deployed in domestic settings shared across family members, potentially reinforcing gendered expectations and exposure [17]. Feminization in agents can reinforce traditional gender scripts around service, support, and submissiveness [9]. Empirical work on gender and SVA privacy has prioritized anthropomorphism, persona design, and interaction style over structural privacy pathways, leaving open questions about how gender informs perceived risk, benefit, trust, control, and protective action in voice-mediated systems [10].

### C. Structural Modeling of Privacy Pathways

Recent studies have proposed and validated structural models that link perceived privacy risk, perceived benefits, transparency and trust, self-efficacy, and privacy-protective behavior in SVA and related digital contexts [7], [15], [16]. These models offer an aggregate view of how attitudes and behaviors interrelate, but leave open whether the same pathways operate similarly across gender. Prior work has documented gendered differences in privacy concerns and protective action [5] but has not examined whether structural privacy pathways are gender-differentiated. This study addresses that gap by applying multi-group PLS-SEM to examine gendered differences in both construct levels and path coefficients.

## III. METHODOLOGY

### A. Survey Design and Model

This study extends our earlier findings from [12], in which a five-construct structural equation model capturing youth privacy-protective behavior toward SVAs was developed and verified. The dataset used in this study is the same as that reported in [12], [13], and the same five constructs, reflective indicators, and measurement scales are retained. The model defines Perceived Privacy Risk (PPR) as the extent to which youth feel vulnerable or at risk when using voice-activated AI assistants[18], [19]. Perceived Privacy Benefits (PPBf) represent the perceived advantages or conveniences gained from using these assistants that can offset privacy concerns [20], [21]. Algorithmic Transparency and Trust (ATT) is the degree to which users believe developers are transparent about data practices, thereby fostering trust [22], [23]. Privacy Self-Efficacy (PSE) refers to users' confidence in their ability to identify, manage, and protect personal information when using SVAs [24], [25]. Finally, Privacy-Protective Behavior (PPB) encompasses the concrete actions taken by youth to safeguard their personal information and limit data collection in these assistants [26], [27]. The structural model, detailing the direct effects among these constructs, is replicated in this current analysis (see Fig. 1). Consistent with our analytical focus, this MGA assesses whether the direct and indirect effects specified in the original model differ by gender. This study received ethics approval from the Vancouver Island University Research Ethics Board (VIU-REB; Ref. #103597), for the consent forms, questionnaires, and related amendments. The survey instruments were adapted from constructs validated in prior studies and tailored to the SVA context. The instruments consist of 4 indicators for each of the five constructs. We measured responses to the items on a 5-point Likert scale ranging from 1 ("Strongly disagree") to 5 ("Strongly agree"). Higher scores indicate higher levels of the underlying constructs. We also collected data on control variables, including age, gender, educational level, and SVA usage frequency.

### B. Participant Recruitment and Demographic

Participants were recruited through multiple channels, including flyers, emails, personal networks, LinkedIn, and through collaboration with several Canadian school districts and Universities to reach our targeted demographics of youth aged 16-24. Participation in the survey was entirely voluntary and anonymous. A monetary incentive was offered to the first 500 survey respondents, with district-specific exceptions where required. A consent form was administered before starting the questionnaire. We conducted online surveys through Microsoft

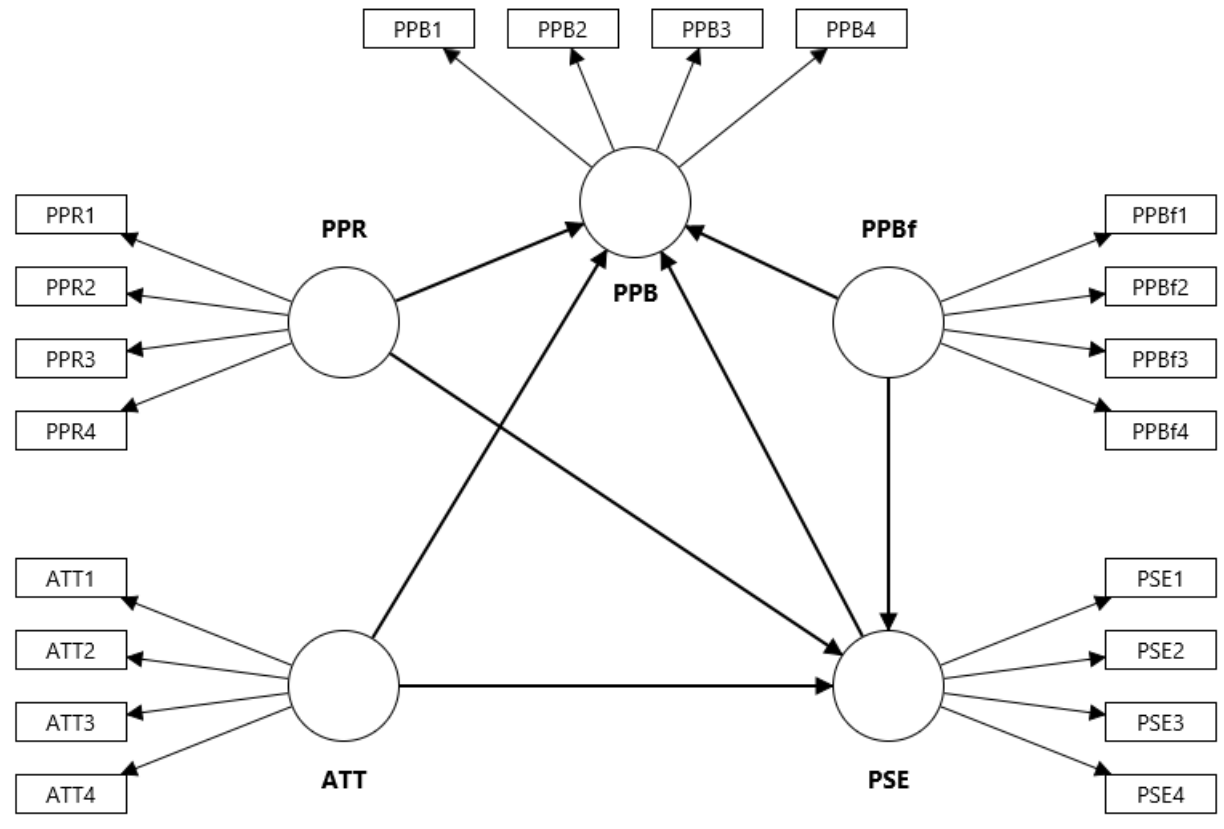

Fig. 1. Conceptual model.

Forms. Upon completing the questionnaire, participants were directed to a separate form to claim the incentive by providing their email address. A total of 494 participants took part in the questionnaire. Responses were omitted that did not meet the demographic criteria (Canadian youths aged 16-24 with at least one SVA use in the prior month) or that contained insufficient data ($\geq$ 20% missing responses). After data cleaning, 469 valid responses were available for analysis. Remaining item-level nonresponses were left blank in the CSV and imported to SmartPLS as missing values (no imputation). SmartPLS handled the remaining missing data with its default pairwise handling during model estimation. The proportion of missing data after cleaning was low ($\leq$ 0.5% for any indicator). Of those 469 valid responses, 174 were classified as female, 241 as male, and 15 as non-binary, and 39 were missing or selected prefer not to say (PNtS). Most participants were in high school (Male: n = 145; Female: n = 99; Non-binary: n = 7; PNtS: n = 26), and the frequency of SVA usage was similar across all groups. Full demographic characteristics are detailed in Table I.

### C. Analysis Procedure

To address RQ1, we computed construct scores as the mean of each construct's indicators (four items per construct), yielding scores on the original 1-5 scale. We compared female and male cohorts using independent-samples t-tests and calculated Cohen's d as the standardized mean difference (SMD), using the *tableone* package in R. Non-binary and "prefer not to say" groups were excluded from inferential tests due to small n. Addressing RQ2, we conducted an MGA in SmartPLS (v 4.1.1.6) to compare female (n=174) and male (n=241) cohorts. After confirming configural invariance, we tested for significant between-group differences in standardized path coefficients using a permutation test (1,000 subsamples, two-tailed $\alpha = 0.05$).

TABLE I. PARTICIPANT DEMOGRAPHIC

| | **Male (n = 241)** | | **Female (n = 174)** | | **Non-binary (n = 15)** | | **Prefer Not to Say (n = 39)** | |
|---|---|---|---|---|---|---|---|---|
| | ***n*** | ***%*** | ***n*** | ***%*** | ***n*** | ***%*** | ***n*** | ***%*** |
| ***Education Level*** | | | | | | | | |
| High School | 145 | 60.2 | 99 | 56.9 | 7 | 46.7 | 26 | 66.7 |
| Post-Secondary | 93 | 38.6 | 75 | 43.1 | 8 | 53.3 | 8 | 20.5 |
| Blank/ Missing | 3 | 1.2 | 0 | 0 | 0 | 0 | 5 | 12.8 |
| ***Frequency of SVA use*** | | | | | | | | |
| Daily | 66 | 27.4 | 45 | 25.9 | 1 | 6.7 | 14 | 35.9 |
| Monthly | 18 | 7.5 | 14 | 8 | 2 | 13.3 | 4 | 10.3 |
| Rarely | 95 | 39.4 | 69 | 39.7 | 8 | 53.3 | 18 | 46.2 |
| Weekly | 61 | 25.3 | 46 | 26.4 | 4 | 26.7 | 2 | 5.1 |
| Blank/ Missing | 1 | 0.4 | 0 | 0 | 0 | 0 | 1 | 2.6 |
| | **Mean (SD)** | | **Mean (SD)** | | **Mean (SD)** | | **Mean (SD)** | |
| ***Age*** | 18.66 (2.38) | | 18.76 (2.31) | | 18.86 (1.70) | | 17.81 (1.65) | |

For all tests, we treat $p < 0.05$ as statistically significant. Findings with $0.05 \leq p < 0.10$ are labeled exploratory and interpreted cautiously. To address RQ3, we examined descriptive patterns for participants who identified as non-binary or selected "prefer not to say," reporting mean construct scores and commenting on notable differences.

## IV. RESULTS

### A. Measurement Model Validation

The constructs used in this study were validated for reliability and validity in a prior study [12], using indicator loadings, Average Variance Extracted (AVE), Composite Reliability (roh_c), Dillon-Goldstein's rho (rho_a), and Heterotrait-Monotrait Ratio (HTMT). All indicator loadings exceeded the recommended threshold of 0.708 [28], with the exception of ATT1 (0.679) and ATT4 (0.702). These items were retained as they remained above the acceptable cutoff of 0.60 [29], and the ATT construct demonstrated satisfactory convergent validity (AVE = 0.529) and composite reliability measures (rho_c = 0.818, rho_a = 0.712). All constructs exceeded the convergent validity threshold of 0.50 and the 0.70 threshold for both reliability metrics [28]. Construct discriminant validity was assessed using the HTMT, with all values falling well below the threshold of 0.85 [28]. For the present gender-based MGA, the measurement model was estimated again separately for the male and female cohorts. The results aligned closely with the prior validation, with only minor deviations in a few indicator loadings and one marginally low AVE value for the ATT construct in the female cohort. As these deviations were minor and did not affect interpretability, detailed group-specific measurement results are summarized rather than reproduced in full; however, values approaching recommended thresholds are reported explicitly to support transparent interpretation in the multigroup setting.

### B. Construct Mean Differences

Mean scores, standard deviations, and between-group comparisons for the male and female cohorts are presented in Table II. On a 5-point Likert scale, mean scores ranged from 2.54 to 3.59 across constructs and genders, with PPR consistently rated highest and ATT lowest. Addressing RQ1, the independent samples t-test and corresponding standardized mean differences (SMD) revealed significant differences between the groups. Males reported significantly higher levels of PSE ($M = 3.08$ vs. $M = 2.84$; $p < 0.01$; SMD = 0.296), representing a small-to-moderate effect size, and marginally higher levels of PPBf ($M = 3.08$ vs. 2.92; $p < 0.1$; SMD = 0.179), reflecting a small effect. This indicates greater confidence in their ability to manage digital privacy and stronger perceived benefits of SVA use. Notably, no other significant differences were noted in ATT ($M = 2.57$ vs. $M = 2.54$; $p > 0.1$; SMD = 0.038), PPR ($M = 3.59$ vs. $M = 3.56$; $p > 0.1$; SMD = 0.030), or PPB ($M = 3.03$ vs. $M = 2.96$; $p > 0.1$; SMD = 0.102), with all SMD values indicating negligible differences. These results suggest that both genders reported similarly lower levels of algorithmic trust and moderate levels of perceived risk in SVA use and protective behaviors.

TABLE II. MEAN DIFFERENCES

| Mean (SD) | Male | Female | p | SMD |
|---|---|---|---|---|
| ATT | 2.57 (0.75) | 2.54 (0.67) | 0.702 | 0.038 |
| PPB | 3.03 (0.78) | 2.96 (0.75) | 0.307 | 0.102 |
| PPBf | 3.08 (0.96) | 2.92 (0.85) | 0.075 | 0.179 |
| PPR | 3.59 (0.96) | 3.56 (0.87) | 0.762 | 0.030 |
| PSE | 3.08 (0.87) | 2.84 (0.74) | 0.004 | 0.296 |

## C. *Measurement Invariance*

The MICOM procedure consists of three steps: (1) configural invariance, (2) composite invariance, and (3) the assessment of equal means and variances. Configural invariance (Step 1) is established by default in SmartPLS [28]. Results for Steps 2 and 3 are presented in Table III.

### 1) *Composite Invariance (Step 2)*

Compositional invariance evaluates whether the composition of the construct is consistent across groups. The test is supported if the original composite correlation meets or exceeds the 5% quantile from the permutation distribution and the corresponding p-value is not significant [28]. All five constructs satisfied these criteria, confirming compositional invariance.

### 2) *Assessment of Equal Means and Variance (Step 3)*

Step 3 evaluates whether the composite scores have equal means (Step 3a) and variances (Step 3b). For full invariance to be established, the mean difference (MD) and variance difference (VD) must fall within the calculated 95% confidence interval, and the associated p-value must be insignificant [30]. Our analysis shows that most constructs have equal means, with the exception of PSE, which failed to demonstrate equal means and variance, indicating that full measurement invariance was not achieved for this construct. These findings align with the significant mean difference observed for PSE. PPBf also demonstrated unequal variance but an equal mean, indicating partial invariance. These results suggest that there are significant differences in how the constructs are perceived or measured across gender groups. Since both PSE and PPBf established compositional invariance in Step 2, meaningful comparisons of their structural path coefficients between groups are still considered permissible [28].

## D. *Multigroup Structural Model*

The MGA was conducted to examine the difference in path coefficients (β). The β values indicate the strength of the relationships between constructs, while the p-values determine their statistical significance. We interpret coefficients in context and report permutation-based p-values for between-group differences [29]. Addressing RQ2, Table IV presents the structural path coefficients for female and male cohorts and the results of the between-group comparisons. The analysis yielded three marginally significant paths. The direct effect of PPR on PPB was stronger for males ($\beta = 0.424$) than for females ($\beta = 0.233$; $p < 0.1$). The indirect effect of ATT on PPB via PSE was stronger for females ($\beta = 0.229$) than for males ($\beta = 0.132$; $p < 0.1$). The indirect effect of PPR on PPB via PSE also differed by gender. For females, the mediation path was positive ($\beta = 0.050$), while for males it was negative ($\beta = -0.016$), resulting in a slightly significant difference ($p < 0.1$).

TABLE III. MEASUREMENT INVARIANCE OF COMPOSITE MODELS

| Construct | Composite (Step 2) | | | |
|---|---|---|---|---|
| | Correlation value | 5% Quintile | p | Composite Invariance |
| ATT | 0.997 | 0.954 | 0.853 | Yes |
| PPB | 0.997 | 0.974 | 0.687 | Yes |
| PPBf | 0.996 | 0.986 | 0.378 | Yes |
| PPR | 0.999 | 0.994 | 0.462 | Yes |
| PSE | 0.999 | 0.994 | 0.626 | Yes |
| | Composite (Step 3a) | | | |
| | MD | 95% CI | p | Equal means |
| ATT | -0.062 | [-0.183, 0.200] | 0.536 | Yes |
| PPB | -0.096 | [-0.197, 0.206] | 0.335 | Yes |
| PPBf | -0.176 | [-0.198, 0.193] | 0.080 | Yes |
| PPR | -0.030 | [-0.184, 0.209] | 0.770 | Yes |
| PSE | -0.276 | [-0.190, 0.182] | 0.002 | No |
| | Composite (Step 3b) | | | |
| | VD | 95% CI | p | Equal variance |
| ATT | -0.236 | [-0.277, 0.289] | 0.114 | Yes |
| PPB | -0.081 | [-0.300, 0.280] | 0.573 | Yes |
| PPBf | -0.297 | [-0.251, 0.243] | 0.013 | No |
| PPR | -0.217 | [-0.275, 0.268] | 0.107 | Yes |
| PSE | -0.309 | [-0.279, 0.237] | 0.021 | No |

TABLE IV. MULTIGROUP ANALYSIS

| Structural Path | β Female | β Male | β Difference | p |
|---|---|---|---|---|
| ATT → PPB | -0.071 | 0.069 | -0.141 | 0.192 |
| PPBf → PPB | -0.125 | -0.111 | -0.015 | 0.893 |
| PPR → PPB | 0.233 | 0.424 | -0.191 | 0.062 |
| PSE → PPB | 0.469 | 0.330 | 0.140 | 0.170 |
| ATT → PSE | 0.489 | 0.400 | 0.089 | 0.308 |
| PPBf → PSE | 0.087 | 0.099 | -0.012 | 0.909 |
| PPR → PSE | 0.106 | -0.049 | 0.155 | 0.138 |
| PPR → PSE → PPB | 0.050 | -0.016 | 0.066 | 0.088 |
| PPBf → PSE → PPB | 0.041 | 0.033 | 0.008 | 0.843 |
| ATT → PSE → PPB | 0.229 | 0.132 | 0.098 | 0.091 |

## E. *Descriptive Patterns for Non-binary and “Prefer Not to Say” Groups*

In addition to the male and female cohorts, the sample included respondents who identified as non-binary ($n = 15$) or selected “prefer not to say” ($n = 39$). Because these groups are comparatively small, we do not conduct inferential testing or structural modeling for them. Instead, we report descriptive construct score patterns to provide initial context for RQ3 and to motivate future work with larger and more inclusive samples capable of supporting multigroup modeling beyond the binary (see Table V).

TABLE V. DESCRIPTIVE STATISTICS

| Mean (SD) | *Non-binary* | *Prefer Not to Say* |
|---|---|---|
| ATT | 1.83 (0.54) | 2.39 (0.70) |
| PPB | 3.32 (0.80) | 3.24 (0.77) |
| PPBf | 2.10 (1.07) | 3.16 (1.02) |
| PPR | 4.22 (0.45) | 3.73 (0.83) |
| PSE | 2.50 (0.80) | 3.05 (0.84) |

Addressing RQ3, several notable patterns emerge. As seen in Table V, non-binary respondents reported the lowest levels of ATT (M = 1.83) and PPBf (M = 2.10) among all gender categories, while also reporting the highest level of PPR (M = 4.22). Conversely, those who responded "Prefer not to say" reported moderate-to-high scores, which were often more aligned with the male and female cohorts, particularly for PSE (M = 3.05) and PPB (M = 3.24). These trends suggest that privacy perceptions and behaviors may vary across the gender spectrum in meaningful ways, highlighting a need for future work with larger, more diverse samples.

## V. DISCUSSION

### A. The Gender Efficacy Gap

Our results reveal a gendered difference in privacy self-efficacy in SVA privacy. Specifically, male youth reported significantly higher levels of PSE than female youth. This gap is further supported through the constructs' measurement structure, as PSE failed to achieve full measurement invariance. This indicates heterogeneity not only in how self-efficacy is expressed, but also in how it may be construed across gender groups. For instance, male respondents may interpret these items more in terms of technical competence, whereas female respondents may understand them more in relation to managing privacy in everyday social settings. Taken together, the evidence points to self-efficacy as an important, and potentially gender-sensitive, lever in youth SVA privacy decision-making. These findings align with prior research on gendered risk perception and confidence, where men often report higher self-efficacy in technology contexts [6]. The failure of PSE to achieve invariance could explain patterns observed in SVA privacy research, where general concerns are high but confidence may be unevenly distributed [8]. The meaning, or perception, of self-efficacy may be gendered, shaped by different socialization experiences with technology [16] and potentially by the feminine, submissive design of voice assistants [9], which could undermine confidence among female users.

### B. Gender as a Moderator in Privacy Decision-Making

MGA revealed that gender moderates the structural pathways connecting perception to actions. Our three marginally significant differences highlight distinct decision-making mechanisms for male and female youth. The direct path from PPR to PPB was stronger for males, aligning with research suggesting men are more likely to translate threats into direct action [6]. In contrast, the indirect effect of ATT on PPB via PSE was stronger for females, suggesting that for young women, trust is a necessary precursor to feeling confident, which is essential for action. This pathway echoes the concept of "online apathy", where a sense of hopelessness around opaque data practices can stifle protective behavior [8], [11]. The nuanced result showing that the path from PPR to PPB through PSE was positive for females but negative for males is notable. For females, higher perceived risk can amplify the mediating role of self-efficacy, suggesting that the perception of threat may necessitate a sense of control as a precursor to protective action, refining findings that risk is a direct deterrent [15]. For males, the relationship differs, indicating that risk perception may influence behavior through more direct routes or that high-risk contexts undermine their sense of efficacy.

### C. Towards Inclusive Privacy Science: Insights from Non-Binary/Other and Prefer-Not-to-Say Participants

The descriptive results for non-binary and "prefer not to say" respondents reveal a profile descriptively different from the binary gender groups, with the lowest ATT, lowest PPBf, and highest PPR. While excluded from formal modeling due to sample size, their results signal a critical finding. Dominant models of privacy calculus, validated on binary gender groups, may fail to capture the lived experiences of gender-diverse youth. While important research has focused on technical and design frameworks for trans and non-binary inclusion in AI systems, outlining best practices and policy recommendations [31], we lack foundational knowledge of privacy psychology for these communities. Therefore, the notable patterns in our data should be an indicator of our need to move beyond the binary and be more inclusive in developing methodological frameworks that include all gender identities.

### D. Implications and Future Work

The results suggest several implications for smart voice assistant design and governance. Privacy self-efficacy appears to be a key lever of protective behavior for all users, especially females. Clear, accessible privacy controls could help bridge the gap between concern and behavior. Marketing strategies should avoid benefit framing that undermines privacy-protective behavior, particularly for men. Finally, platforms and policymakers should consider how SVA privacy controls and documentation can better support non-binary and gender-diverse youth, whose needs current interfaces may overlook. This study has some limitations to acknowledge. The cross-sectional design, reliance on self-reported behavior, and a Canadian sample limit cross-cultural generalizability. The primary limitation is the exclusion of non-binary and other gender-diverse identities from the primary MGA due to insufficient statistical power. While our descriptive data for these groups reveal descriptively different privacy profiles, we cannot model their structural pathways. Future research requires a longitudinal design, cross-cultural replication, and recruitment to build models that move beyond the gender binary.

## VI. CONCLUSION

This study investigated gender-differentiated pathways to privacy protection among youth using SVAs. Our findings reveal a gender efficacy gap, with male youth reporting significantly higher PSE than females. While both genders showed similar levels of PPR and PPB, the structural pathway linking perception to action was moderated by gender. For males, the direct path from PPR to PPB was stronger, suggesting a more straightforward translation of risk into action. For females, ATT played a more important indirect role, strengthening self-efficacy before leading to protective behavior. Additionally, descriptive insight from non-binary and undisclosed gender respondents demonstrated distinct privacy profiles, including lower levels of ATT and PPBf, and higher levels of PPR, highlighting the need for more inclusive privacy research. These findings extend prior models by demonstrating that gender shapes privacy perceptions and pathways to protections. The results emphasize that fostering trust may be especially important for empowering young women, while risk communication might more directly motivate young men. Privacy support in SVAs must move beyond a one-size-fits-all approach and toward tailored interventions and design, fostering confident, capable digital citizens across all gender identities.

## Acknowledgment

This project has been funded by the Office of the Privacy Commissioner of Canada (OPC); the views expressed are those of the authors and do not necessarily reflect those of the OPC.